\documentstyle[12pt]{article}
\input psfig.sty
\def\nn{\noindent}
\def\Re{{\cal R \mskip-4mu \lower.1ex \hbox{\it e}\,}}
\def\Im{{\cal I \mskip-5mu \lower.1ex \hbox{\it m}\,}}
\def\ie{{\it i.e.}}
\def\eg{{\it e.g.}}
\def\etc{{\it etc}}
\def\etal{{\it et al.}}

\def\sub#1{_{\lower.25ex\hbox{$\scriptstyle#1$}}}
\def\tev{\,{\ifmmode\mathrm {TeV}\else TeV\fi}}
\def\gev{\,{\ifmmode\mathrm {GeV}\else GeV\fi}}
\def\mev{\,{\ifmmode\mathrm {MeV}\else MeV\fi}}
\def\to{\rightarrow}

\def\subw{_{\rm w}}
\def\mh{\ifmmode m\sbl H \else $m\sbl H$\fi}
\def\mch{\ifmmode m_{H^\pm} \else $m_{H^\pm}$\fi}
\def\mt{\ifmmode m_t\else $m_t$\fi}
\def\mc{\ifmmode m_c\else $m_c$\fi}
\def\mz{\ifmmode M_Z\else $M_Z$\fi}
\def\mw{\ifmmode M_W\else $M_W$\fi}
\def\mws{\ifmmode M_W^2 \else $M_W^2$\fi}
\def\mhs{\ifmmode m_H^2 \else $m_H^2$\fi}   
\def\mzs{\ifmmode M_Z^2 \else $M_Z^2$\fi}
\def\mts{\ifmmode m_t^2 \else $m_t^2$\fi}
\def\mcs{\ifmmode m_c^2 \else $m_c^2$\fi}
\def\mchs{\ifmmode m_{H^\pm}^2 \else $m_{H^\pm}^2$\fi}
\def\ztwo{\ifmmode Z_2\else $Z_2$\fi}
\def\zone{\ifmmode Z_1\else $Z_1$\fi}
\def\mtwo{\ifmmode M_2\else $M_2$\fi}
\def\mone{\ifmmode M_1\else $M_1$\fi}
\def\tb{\ifmmode \tan\beta \else $\tan\beta$\fi}
\def\xw{\ifmmode x\subw\else $x\subw$\fi}
\def\ch{\ifmmode H^\pm \else $H^\pm$\fi}
\def\lum{\ifmmode {\cal L}\else ${\cal L}$\fi}
\def\inpb{\,{\ifmmode {\mathrm {pb}}^{-1}\else ${\mathrm {pb}}^{-1}$\fi}}
\def\infb{\,{\ifmmode {\mathrm {fb}}^{-1}\else ${\mathrm {fb}}^{-1}$\fi}}
\def\epem{\ifmmode e^+e^-\else $e^+e^-$\fi}
\def\ppb{\ifmmode \bar pp\else $\bar pp$\fi}
\def\bsg{\ifmmode B\to X_s\gamma\else $B\to X_s\gamma$\fi}
\def\bsll{\ifmmode B\to X_s\ell^+\ell^-\else $B\to X_s\ell^+\ell^-$\fi}
\def\bstt{\ifmmode B\to X_s\tau^+\tau^-\else $B\to X_s\tau^+\tau^-$\fi}
\def\lamt{\ifmmode \tilde\lambda\else $\tilde\lambda$\fi}
\def\shat{\ifmmode \hat s\else $\hat s$\fi}
\def\that{\ifmmode \hat t\else $\hat t$\fi}
\def\uhat{\ifmmode \hat u\else $\hat u$\fi}

\newskip\zatskip \zatskip=0pt plus0pt minus0pt
\def\matth{\mathsurround=0pt}

\def\atversim#1#2{\lower0.7ex\vbox{\baselineskip\zatskip\lineskip\zatskip
  \lineskiplimit 0pt\ialign{$\matth#1\hfil##\hfil$\crcr#2\crcr\sim\crcr}}}

\renewcommand{\thefootnote}{\fnsymbol{footnote}}

\begin{document} \begin{titlepage} 
\rightline{\vbox{\halign{&#\hfil\cr
&SLAC-PUB-7702\cr
&December 1997\cr}}}
\begin{center}

{\Large\bf Failure of JoAnne's Global Fit to the Wilson Coefficients in 
Rare B Decays: A Left-Right Model Example}
\footnote{Work supported by the Department of 
Energy, Contract DE-AC03-76SF00515}
\medskip

\normalsize 
{\large Thomas G. Rizzo } \\
\vskip .3cm
Stanford Linear Accelerator Center \\
Stanford CA 94309, USA\\
\vskip .3cm

\end{center}

\begin{abstract} 

In the Standard Model and many of its extensions, it is well known that all 
of the observables associated with the rare decays $b\to s\gamma$ and 
$b\to s\ell^+\ell^-$ can be expressed in terms of the three Wilson 
coefficients, $C_{7L,9L,10L}(\mu \sim m_b)$, together with several universal 
kinematic functions. In particular it has been shown that the numerical values 
of these coefficients can be uniquely extracted by a three parameter global 
fit to data obtainable at future $B$-factories given sufficient integrated 
luminosity. In this paper we examine if such global fits are also 
sensitive to new operators beyond those which correspond to the above 
coefficients, \ie, whether is it possible that new operators can be of 
sufficient importance for the three parameter fit to fail and for this 
to be experimentally observable. Using the Left-Right Symmetric Model 
as an example of a scenario with an extended operator basis, we demonstrate 
via Monte Carlo techniques that such a possibility can indeed be realized. 
In some sense this potential failure of the global fit approach can actually 
be one of its greatest successes in identifying the existence of new physics.

\end{abstract} 




\renewcommand{\thefootnote}{\arabic{footnote}} \end{titlepage}


\section{Introduction}

Rare decays of heavy quarks which do not occur at the tree 
level in the Standard Model (SM) can provide a unique opportunity for new 
physics to reveal itself.  
When such decays occur through loops then the participating SM particles and 
those associated with the new interaction are placed on the same 
footing and may yield comparable contributions to the various decay 
amplitudes. In these cases it may be possible to 
isolate such additional contributions and learn something about the detailed 
nature of the new physics scenario. 

Amongst the rare decays involving $b$ 
quarks, two of the cleanest and most well understood inclusive processes are 
$b\to s \gamma$ and $b\to s\ell^+\ell^-$. At present, the branching fraction 
for the $b\to s\gamma$ mode has been measured by CLEO{\cite {cleo}} to be 
$B(b\to s\gamma)=(2.32\pm 0.57 \pm 0.35)\times 10^{-4}$ while a preliminary 
result from ALEPH{\cite {aleph}} yields the value 
$(3.38\pm 0.74 \pm 0.85)\times 10^{-4}$. On the other hand, there exist only 
upper bounds for the decay $b\to s\ell^+\ell^-$; the strongest constraint 
at present is the $90\%$ CL limit $B(b\to s\ell^+\ell^-)<4.2\times 
10^{-5}${\cite {cleo2}} from CLEO, which is obtained by 
combining both their di-electron and di-muon data samples. As we will see 
below, this is only a factor of $\sim 6$ above the expectations of the SM for 
this branching fraction so that we may expect this decay to be observed in 
the near future. 

In the SM and in many of its extensions (including, \eg, fourth generation 
models, models with an extra down-type quark, SUSY, extended 
Higgs sectors, $Z'$ scenarios, theories 
with large anomalous gauge boson couplings, \etc.) the phenomenology of 
both of these rare processes above are 
almost completely determined by the numerical values of the Wilson 
coefficients of only a small set of operators evaluated at 
the scale $\mu \sim m_b$. In our somewhat unconventional notation these are 
denoted as $C_{7L,9L,10L}(\mu)$. At the weak scale the operators corresponding 
to these coefficients arise in the SM from the usual $\gamma$ and $Z$ 
penguins as well as $W$ box diagrams. It has been successfully argued in the 
literature{\cite {jlh}} that by combining observables associated with 
both the $b\to s \gamma$ and $b\to s\ell^+\ell^-$ processes, a model 
independent three dimensional global fit can be performed to numerically
determine the values of these three Wilson coefficients. Indeed, given 
sufficient statistics at future $B$-factories this 
approach leads to only rather modest uncertainties in the fitted values of 
these coefficients allowing us to test the SM and look for new physics.
We note that only the magnitude $|C_{7L}(\mu)|$ can be extracted from 
the $b\to s\gamma$ transition so 
that its sign would remain undetermined from this channel alone even if 
infinite precision were available. For the observables associated with the 
$b\to s\ell^+\ell^-$ decay all three of the coefficients contribute and 
therefore their 
relative signs as well as their magnitudes can be extracted from the data when 
combined with our knowledge of $B(b\to s\gamma)$. 

In some ways the determination of the these three Wilson coefficients via a 
global fitting procedure in rare $B$ decays is similar to the searches for 
new physics in precision electroweak
measurements{\cite{obl}} through the use of the oblique 
parameters $S,T,U${\cite {joa}}. As the reader may recall, in the SM (for a 
reference value of the top quark and Higgs boson masses) these parameters are 
all identically zero. For certain classes of new physics, such as a fourth 
generation of quarks and leptons, 
fits to precision data would then lead to some consistent set of non-zero 
values for these parameters with a good $\chi^2$. Of course, if 
these parameters are found to be non-zero and it is also found 
that different precision 
observables yielded statistically distinct $S,T,U$ values then we would 
necessarily conclude that the new interactions are not describable by the 
oblique corrections alone. (As is well known, any class of new interactions
that induce significant flavor dependent vertex corrections, 
such as a $Z'$, can not be portrayed solely in terms of $S,T,U$.) Such a 
situation would provide a unique window on the complex nature of the new physics
scenario.

It is then obvious that when sufficient statistics become available in the 
not too distant future for this type of analysis of rare $B$ decays to be 
performed,  there are only three possible 
outcomes: ($i$) The numerical values for the coefficients 
are found to agree with the SM expectations with a good $\chi^2$. In this case 
the new physics is decoupled 
and either higher precision data or searches elsewhere are necessary to 
uncover it. ($ii$) A quality fit is obtained but the values one finds for the 
three Wilson coefficients are far from the SM expectations in $\chi^2$. This 
is the result usually discussed and anticipated in the 
literature{\cite {jlh}} in the set of extended models listed 
above. ($iii$) As with the case of precision measurements, the 
last possibility is potentially the most interesting and the one we are 
interested in here: the value of $\chi^2$ for the best three parameter fit is 
found to be very large and cannot be accounted for by an under estimation of 
systematic uncertainties. This represents in some sense a failure of the model 
independent approach in that it is clear that the true numerical values of 
the three Wilson coefficients are not being extracted from the data. However, 
another point of view is that this result is in fact this approach's greatest 
triumph since it is telling us that the new interactions necessarily involve 
an extension of the operator basis to include {\it new} operators beyond the 
usual set. This implies that the new physics 
scenario is richer than any of those in the list above. 

Thus the question we wish to address here is whether new physics which does
involve new operators in an extended basis can indeed manifest itself as a 
poor fit when we have the freedom to vary the 3 coefficients to 
obtain a good $\chi^2$. (Of course, $|C_{7L}(\mu)|$ cannot be freely varied 
by too large an amount due to the reasonable 
agreement between the present data and SM expectations for the $b\to s\gamma$ 
decay rate as discussed below.) The purpose of this 
paper is then to demonstrate this result by providing an existence proof that 
a new physics scenario of the desired type not only exists, in the form of the 
Left-Right Symmetric Model (LRM), but that it leads to poor values of
$\chi^2$ in the global fit when only the usual three 
operators are employed. The point we wish to stress here is not the 
particular physics of the LRM, or any other specific model, but that the 
existence of an extended operator 
basis can indeed manifest itself in the poorness of the three parameter 
fit given reasonable integrated luminosities. We note, however, that without 
further analysis the failure of the fit itself will not yield information on
which new operators would need to be introduced.  We 
further note that the LRM is of course not the only new physics scenario 
with an extended operator basis{\cite {two}}.

The outline of this paper is as follows. In Section 2 we overview the status 
of the various pieces necessary for calculations of the $b\to s\gamma$ and 
$b\to s\ell^+\ell^-$ decay rates and distributions in the SM. In Section 3 
we provide a background on the basics of the LRM and the parameters it contains 
which are relevant for the processes of interest here. We discuss and 
generate several possible forms of the right-handed CKM weak mixing matrix, 
$V_R$, following a systematic approach that 
maintains unitarity and wherein agreement with experimental constraints can 
be easily accommodated. Next we set up the LRM calculations for these decays 
and demonstrate that many LRM parameter space regions exist wherein the rate 
for $b\to s\gamma$ is essentially the
same as in the SM. The general formulae for the 
$b\to s\ell^+\ell^-$ double differential distributions are discussed and 
it is shown that the LRM yields distinct results for observables associated 
with this decay 
even when the $b\to s\gamma$ rate duplicates the SM expectation.  In 
Section 4 we discuss our Monte Carlo approach and generate a large number 
of data samples corresponding to each of the several models discussed in 
the previous section. We demonstrate that for 
high luminosities, corresponding to $5\times 10^8$ $B\bar B$ pairs, typical 
of samples available at hadron colliders, the resulting fits to the 
conventional three Wilson coefficients lead to very large $\chi^2$ values 
which clearly signal the failure of the fit. For lower luminosities, now 
corresponding to $5\times 10^7$ $B\bar B$ pairs, typical of samples to be 
available at $\Upsilon$(4S) machines, the $\chi^2$ values are also found to 
be quite large in 
most, but not all, cases. A discussion of these results and our conclusions 
can be found in Section 5.

\section{Rare Decays in the Standard Model}

In order to be convinced that new physics has indeed been discovered it is 
necessary that the predictions for the rates and other observables associated 
with these decays in the SM be on firm ground. In particular, calculations of 
the decay rates and, 
in the case of $b\to s\ell^+\ell^-$, kinematic distributions have become 
increasingly sophisticated within the SM context. A ``straightforward'' 
next-to-leading order(NLO) calculation finds 
$B(b\to s\gamma)=(3.28\pm 0.33)\times 10^{-4}${\cite {bsg}}. However the 
inclusion of $1/m_b^2$ and $1/m_c^2$ corrections{\cite {corr}} increases this 
value by about $3\%$. A further enhancement of about $3\%$ occurs 
when one systematically disregards any of the NNLO terms{\cite {buras}}. While 
closer to the central value of the preliminary ALEPH measurement these 
predictions are certainly consistent with the present CLEO data.  For purposes
of simplicity, in our numerical analysis below 
we make direct use the NLO result and ignore these additional small 
correction terms. This approximation will have no impact on our conclusions. 
One may anticipate that in 
the next few years this theoretical uncertainty may shrink to as low as $5\%$ 
as the various input parameters are better determined. A comparable 
experimental determination of this branching fraction may also be possible at 
future $B$-factories since the measurements will be limited only by 
systematics. We will {\it assume} below that the SM value for this 
branching fraction is essentially realized by future experiments within 
the experimental and theoretical uncertainties.

In the case of 
$b\to s\ell^+\ell^-$, a complete short-distance NLO calculation has been 
available for some time{\cite {bsll}} and the $1/m_b^2$ correction terms are 
also known{\cite {corr2}} but provide only very small modifications. One of the 
remaining difficulties is the inclusion 
of non-perturbative long distance pieces associated with the 
$J/\psi$ and $\psi'$ resonances and the corresponding $1/m_c^2$ corrections.
Here some modeling uncertainties remain and the traditional approach 
has been to treat the resonance contributions phenomenologically, which is not 
without some difficulties{\cite {corr3}}. However, at least in the regions 
sufficiently below and above the resonances (\ie, $s=q^2/m_b^2 \leq 0.3$ or 
$\geq 0.6$, where $q^2$ is the invariant mass of the lepton pair)
Buchalla, Isidori and Rey{\cite {corr}} have shown that the 
heavy quark expansion in terms of $1/m_c^2$ leads to reliable predictions with 
only small corrections to the NLO results and that there are
no difficulties associated with double counting. In the same kinematic regions 
the phenomenological resonance models give comparable numerical predictions. 
In many extensions to the SM, the $1/m_{c,b}^2$ and resonant contributions are 
either of the same form as in the SM or can be easily obtained via suitable 
modifications of the SM terms and we employ these results in our analysis 
below. We note that the SM predicts a branching fraction for 
$b\to s \mu^+\mu^-$ of 
$\simeq 6\times 10^{-6}$, which is not too far from the present upper bound. 

In order to obtain the complete parton level NLO predictions for these two 
processes in the SM (or in other models), several steps are necessary. First, 
the complete operator basis must be determined at the high (matching) scale, 
typically taken to be $M_W$. 
Secondly, the matching conditions for the coefficients of the operators at 
the high scale must be calculated at both the LO and NLO. Thirdly, the 
anomalous dimension matrices for the relevant operators at both LO and NLO 
are determined and the coefficients are evolved to the $\mu \sim m_b$ scale 
via the Renormalization Group Equations. 
Lastly, at the scale $\mu$ the matrix elements of the relevant 
operators need to be computed through NLO. For the SM all of these pieces are  
now essentially in place for both the $b\to s\gamma$ and $b\to s\ell^+\ell^-$ 
decays after an enormous amount of labor. Unfortunately, the corresponding 
results only partially exist for most of these pieces in almost all 
extensions to the SM{\cite {higg}}.

\section{The Left-Right Model}

\subsection{Model Background}

In order to be self-contained we briefly review the relevant parts of the 
LRM needed for the discussion below; for details of the model the reader is 
referred to Ref. {\cite {moha}}. The LRM is based on the 
extended gauge group $SU(2)_L \times SU(2)_R \times U(1)$ and can lead to 
interesting new effects in the B system{\cite {vol}}. Due to the extended 
gauge structure 
there are both new neutral and charged gauge bosons, $Z'$ and $W^{\pm}_R$, in 
addition to those present in the Standard Model. In this scenario the 
left-(right-)handed (LH, RH) fermions of the SM are assigned to doublets under 
the $SU(2)_{L(R)}$ group and a RH neutrino is introduced. The Higgs fields 
which can directly 
generate SM fermion masses are thus in bi-doublet
representations, \ie, they transform as doublets under both $SU(2)$ groups. 
The LRM is quite robust 
and possesses a large number of free parameters which play an interdependent 
role in the calculation of observables and in obtaining the existing 
constraints on the model resulting from various experiments. 

As far as B physics and the subsequent discussion are 
concerned there are several parameters of direct interest, most of which result
from the structure and spontaneous symmetry breaking of the extended gauge 
sector.  The ratio of the $SU(2)_R$ and $SU(2)_L$ gauge couplings is bounded
by $0.55<\kappa=g_R/g_L\leq 2$, where the lower limit is a model constraint 
and the upper one is simply a naturalness assumption. 
Whereas $g_L$ is directly related to $e$ as usual through $\sin^2 \theta_W$, 
$g_R$ is unconstrained except through the definition of electric charge and 
naturalness arguments; GUT embedding scenarios generally suggest that 
$\kappa \leq 1$. For simplicity we assume $\kappa=1$ in most of our 
discussion below.  The $SU(2)_L \times SU(2)_R\times U(1)$ extended symmetry 
is broken down to the SM 
via the action of Higgs fields that transform either as doublets as discussed
above, or also possibly as triplets 
under $SU(2)_R$. This choice of Higgs representation determines both the mass 
relationship 
between the $Z'$ and $W_R$ (analogous to the condition that $\rho=1$ in the 
SM) as well as the nature of neutrino masses.  In particular, the Higgs triplet 
choice allows for the implementation of the see-saw mechanism and yields a 
heavy RH neutrino. We assume triplet breaking below so that the $Z'$ mass is 
completely specified by the $W_R$ mass and the value of $\kappa$.

After complete symmetry breaking the $W_L$ and $W_R$ bosons mix; this mixing
being described 
by two parameters, a real mixing angle $\phi$ and a phase $\omega$. Note 
that it is usually $t_\phi=\tan \phi$ which appears in expressions directly 
related to observables. The additional phase, as always, can be a new source 
of CP violation. However, in discussing processes in which the RH neutrinos 
do not participate, as is the case in B decays, this angle can be thought of 
as an overall phase of the right-handed 
CKM matrix, $V_R$, and we 
will subsequently ignore it. The mixing between $W_L$ and $W_R$ results in 
the mass eigenstates $W_{1,2}$, with a ratio of masses $r=M_1^2/M_2^2$
(with $M_2 \simeq M_R$). In most models $t_\phi$ is then 
naturally of order a few 
times $r$, or less, in the large $M_2$ limit. Of course, $W_1$ is the state 
directly being produced at 
both the Tevatron and LEPII and is identical to the SM $W$ in the limit
$\phi \to 0$.
We note that when $\phi$ is non-zero, $W_1$ no longer couples to a 
purely LH current. Of course if a heavy RH neutrino is indeed realized then 
the effective leptonic current coupling to $W_1$ remains purely LH as 
far as low energy experiments are concerned. 
As is well-known, one of the strongest classical constraints on this model 
arises from polarized $\mu$ decay{\cite {muon}}, which is trivial to satisfy 
in this case. 

It is important to recall that the extended Higgs sector 
associated with both the breaking of the full LRM gauge group down to 
$U(1)_{em}$ and with the complete generation of fermion masses may also play an 
important role in low energy physics through the existence of 
complex Yukawa and/or flavor-changing neutral current type couplings. However, 
this sector of the LRM 
is highly model dependent and is of course quite sensitive to the detailed 
nature of the fermion mass generation problem. For purposes of brevity and 
simplicity these too will be ignored in the following discussion and we will 
focus solely on the effects associated with $W_{1,2}$ exchange. 

Additional parameters arise in the quark sector.  In principle the effective 
mass matrices for the SM fermions may be non-hermitian implying that the two 
matrices involved in the bi-unitary transformation needed to diagonalize them 
will be unrelated. This means that the elements of the mixing matrix
$V_R$ appearing in the RH charged current for quarks will be unrelated
to the corresponding elements of $V_L=V_{CKM}$. $V_R$ will then involve 3 new 
angles as well as 6 additional phases all of which are {\it a priori} 
unknown parameters. Needless to say the additional phases can be a further 
source of CP violation. 
The possibility that $V_L$ and $V_R$ may be unrelated is 
often overlooked when considering the potential impact of the LRM on low 
energy physics and there has been very little detailed exploration of this 
more general 
situation. Clearly as the elements of $V_R$ are allowed to vary the 
impact of the extended gauge sector on B physics will be greatly affected. 
Other well-known constraints on the LRM, such as universality, the apparent 
observed unitarity of the CKM matrix, $B^0-\bar B^0$ mixing, 
the $K_L-K_S$ mass difference{\cite {bbs}}, as well as Tevatron direct $W'$ 
searches{\cite {cdfd0}}, are quite 
sensitive to variations in $V_R${\cite {oldt}}, but $W_2$ masses as low as 
$450-500$ GeV can be easily accommodated by the present data. To be safe and 
to keep future $W_2$ searches in mind, however, we will generally assume that 
$M_2 \geq 800$ GeV for any form of $V_R$, implying that the magnitude of 
$t_\phi$ is less than $a~few \cdot 10^{-2}$. In our numerical study below we 
will assume various different forms for $V_R$; in all cases we will assume for 
simplicity that the values of the elements of $V_L$ as extracted by current 
experiment{\cite {rev}} are not much influenced by the existence of the new 
LRM interactions. An updated analysis on the possible general structure of 
$V_R$ has yet to be performed.

\subsection{Forms of $V_R$}

In order to study the potential influence of the LRM on $B$ physics various 
forms of $V_R$ should be examined. In fact, the possibility that $V_L\neq V_R$ 
and that $B$ decays are {\it purely} right-handed was entertained some time ago 
by Gronau and Wakaizumi{\cite {gron}}. Just how large the right-handed 
contribution to $b\to c$ transitions is allowed to be within the LRM context is 
not yet accurately known{\cite {me2}} but may be sizeable in magnitude 
with an unknown relative phase. The experimental bounds are as follows.  L3 
has compared\cite{l3} their measurements of both the lepton and missing energy 
spectra in semileptonic $b\to c$ decays with a number of 
different hypotheses, and have excluded both the $(V+A)\times (V-A)$ and 
$V\times (V-A)$ scenarios, clearly indicating that this coupling is dominantly 
left-handed. This qualitative result is confirmed by the sign of the 
$\Lambda_b$ polarization observed by ALEPH{\cite {aleph2}} at the $Z$ pole. 
The strongest constraint comes from CLEO{\cite {cleo3}}, with measurements of
both the 
leptonic forward-backward asymmetry as well as the $D^*$ polarization in the 
decay $B\to D^*\ell \nu$. 
On the theoretical side, Voloshin{\cite {vol}} has recently 
considered how significant right-handed $b\to c$ currents, at the $\sim 15\%$ 
level, may assist in our 
understanding of the $B$ semileptonic branching fraction. 
In a more general context, other forms 
of $V_R$ have been discussed by Langacker and Sankar{\cite {lang}} upon which 
we generalize in the following analysis. 

Since we are not concerned with 
$CP$ violation in what follows, for numerical simplicity we will ignore the 
phases in this matrix in which case it can be completely described by three 
mixing angles. Even in this limiting scenario the set of possible forms for 
$V_R$ is enormous; however, it is sufficient for our purposes here
to simply examine some sample forms. We will assume that each row 
and column of $V_R$ contains only one large element with a magnitude near 
unity as is the true for the conventional 
CKM matrix. In this case there are only six 
matrices about which we can perturb; we write these symbolically as 
\begin{eqnarray}
{\cal M}_A & =  \left( \begin{array}{ccc}
                         1 & 0 & 0 \\
                         0 & 1 & 0 \\
                         0 & 0 & 1
                         \end{array}\right)\,,\quad\quad
{\cal M}_D & =  \left( \begin{array}{ccc}
                         0 & 1 & 0 \\
                         0 & 0 & 1 \\
                         1 & 0 & 0
                         \end{array}\right)\,, \nonumber\\
{\cal M}_B & =  \left( \begin{array}{ccc}
                         0 & 1 & 0 \\
                         1 & 0 & 0 \\
                         0 & 0 & 1
                         \end{array}\right)\,,\quad\quad
{\cal M}_E & =  \left( \begin{array}{ccc}
                         0 & 0 & 1 \\
                         1 & 0 & 0 \\
                         0 & 1 & 0
                         \end{array}\right)\,, \\
{\cal M}_C & =  \left( \begin{array}{ccc}
                         1 & 0 & 0 \\
                         0 & 0 & 1 \\
                         0 & 1 & 0
                         \end{array}\right)\,,\quad\quad
{\cal M}_F & =  \left( \begin{array}{ccc}
                         0 & 0 & 1 \\
                         0 & 1 & 0 \\
                         1 & 0 & 0
                         \end{array}\right)\,. \nonumber
\end{eqnarray}
For example, matrix D corresponds, in the standard CKM 
parameterization{\cite {pdg}}, to the situation  
where $s_{12}, s_{23}$ and $c_{13}$ are all $\sim 1$. Following 
Wolfenstein{\cite {wolf}}, this suggests taking $c_{12}\sim \lambda^n$, 
$s_{13}\sim A\lambda^m$, and $c_{23}\sim B\lambda^p$, where $A,B$ are order 
unity, $n,m,p\geq 1$ and $\lambda \simeq 0.22$. Of course, all these 
parameters are not arbitrary since the experimental constraints from $K_L-K_S$ 
and $B-\bar B$ mixing must be satisfied. For $r\leq 10^{-2}$ and 
$|t_\phi| \leq a~few \times 10^{-2}$, these bounds are trivially fulfilled,
without further fine tuning, if $p\geq 3$ and $n,m\geq 1$. For 
$p\geq 3$, $B$ physics is not very sensitive to the value of $m$ so we take 
for simplicity $p=3$ and $m=1$. This leaves us with a set of matrices we can 
label as $V_R=D(n)$, for n=1,2,3,.... Table 1 lists the 
complete set of parameterizations for all six types of $V_R$ mixing matrices,
which we arrive at by following a similar procedure,
and the constrained range of the exponents we use in the
analysis below. The values of these powers reflect simplicity as well as that 
required to satisfy the low energy experimental constraints. 
%
\begin{table*}[htpb]
\leavevmode
\begin{center}
\label{$V_R$ matrices}
\begin{tabular}{lccccccc}
\hline
\hline
Matrix &$s_{12}$&$c_{12}$&$s_{13}$&$c_{13}$&$s_{23}$&$c_{23}$&constraints \\
\hline
A &$\lambda^n$ & 1 & $B\lambda^p$ & 1 & $A\lambda^m$ & 1 &$n=2,p=3,m=1,2,3$\\
B &1 & $\lambda^n$ & $B\lambda^p$ & 1 & $A\lambda^m$ & 1 &$n=2,m=3,p=1,2,3$\\
C &$A\lambda^m$ & 1 & $B\lambda^p$ & 1 & 1 & $\lambda^n$ &$p=1,m=2,n=1,2,3$\\
D &1 & $\lambda^n$ & $A\lambda^m$ & 1 & 1 & $B\lambda^p$ &$p=3,m=1,n=1,2,3$\\
E &$A\lambda^m$ & 1 & 1 & $\lambda^n$ & 1 & $B\lambda^p$ &$p,m\geq 2,n\geq 1$\\
F &$A\lambda^m$ & 1 & 1 & $\lambda^n$ & $B\lambda^p$ & 1 &$p,m\geq 2,n\geq 4$\\
\hline
\hline
\end{tabular}
\caption{Parameterizations of $V_R$ in the absence of $CP$ violation assuming 
$\lambda \sim 0.2$, $A$ and $B$ are of order unity, and $m,n,p \geq 1$.}
\end{center}
\end{table*}

Note that we have not employed any constraint on the allowed strength of 
the right-handed $b\to c$ coupling in our discussion. 
Since these bounds are relatively complicated their detailed numerical 
impact will be discussed elsewhere{\cite {me2}}, 
but will have the most impact in the cases of matrices $C$ and $D$.

\subsection{Rare Decay Formalism}

The analysis of the decays $b\to s\gamma$ and  $b\to s\ell^+\ell^-$  in the 
LRM begins with the following extended effective Hamiltonian, 
\begin{equation}
{\cal H}_{eff} = {4G_F\over\sqrt 2}\sum_{i=1}^{12}C_{iL}(\mu)
{\cal O}_{iL}(\mu)+L\to R \,,
\end{equation}
The ${\cal O}_{iL,R}$ are the complete set of operators involving only the 
light fields which govern $b\to s$ transitions. The first thing to notice is 
that for convenience the conventional CKM factors in front have been absorbed 
into the values of the coefficients themselves. This is generally a useful 
approach to follow when multiple mixing angle structures appear simultaneously. 
The second point to note 
here is that whereas there are only 10 local operators 
describing $b\to s$ transitions in the SM, \ie, ${\cal O}_{1L}-
{\cal O}_{10L}$, here there are 24 operators. 10 of the additional operators 
correspond to the chiral partners, with $L\to R$, of those present in the 
SM. The complete basis for each helicity structure then 
consists of the usual six 4-quark operators ${\cal O}_{1-6L,R}$, 
the penguin-induced electro- and chromo-magnetic operators respectively 
denoted as ${\cal O}_{7,8L,R}$, as well as 
${\cal O}_{9L,R}\sim e\bar s_{L,R\alpha}\gamma_\mu b_{L,R\alpha}\bar\ell
\gamma^\mu\ell$, and ${\cal O}_{10L,R}\sim e\bar s_{L,R\alpha}\gamma_\mu 
b_{L,R\alpha}\bar\ell\gamma^\mu \gamma_5\ell$ which arise from box diagrams 
and electroweak penguins. (Here the indices $\alpha$ label the color structure 
of the operators.)  In the LRM we not only have the 
augmentation of the operator basis via the obvious doubling of $L\to R$, 
but two new additional 4-quark 
operators of each helicity structure are also present at the tree-level due to 
a possible mixing between the $W_{L,R}$ gauge bosons:  
${\cal O}_{11L,R}\sim (\bar s_\alpha \gamma_\mu c_\beta)_{R,L}
(\bar c_\beta \gamma^\mu b_\alpha)_{L,R}$ and ${\cal O}_{12L,R}\sim 
(\bar s_\alpha \gamma_\mu c_\alpha)_{R,L}(\bar c_\beta \gamma^\mu 
b_\beta)_{L,R}$. Operators ${\cal O}_{12L,R}$ occur at the tree level 
in a fashion analogous to operators ${\cal O}_{2L,R}$. 

In evolving down from the weak scale to $\mu \sim m_b$ these operators mix 
under renormalization as usual. Fortunately the $24\times 24$ anomalous 
dimension matrices split into two identical $12\times 12$ chiral submatrices 
since the operators of each chirality do not 
mix under RGE evolution. The complete $12\times 12$ anomalous dimension matrix 
at LO was first calculated by Cho and Misiak{\cite {old}} and at NLO only the 
$10\times 10$ submatrix corresponding to the SM operators is presently known.

The determination of the matching conditions for the 24 operators at the 
electroweak scale even at LO is already 
somewhat cumbersome since the LRM contains a large number of 
parameters and, in addition to new tree graphs, 116 one-loop graphs must also 
be calculated. (Additional diagrams due to possible physical Higgs exchange are 
not yet included and would introduce additional model dependence.) Some of 
these diagrams have already been calculated for the
earlier analyses of the decay $b\to s\gamma$ in the LRM{\cite {old}}. 
Clearly, NLO matching 
conditions do not yet exist for this model so we employ only the LO ones 
below. However we note that the numerical 
size of these NLO contributions in the SM case were found to be small. 
In addition one has to separately include the new LRM contributions to the 
semileptonic branching fraction, $B_l$, including both finite 
$m_c/m_b\simeq0.29$ 
and LO QCD corrections, since it is conventionally used to normalize 
both the $b\to s\gamma$ and $b\to s\ell^+ \ell^-$ decay rates. 
(We assume that the relevant elements of $V_L$ have the same 
numerical values in the LRM as in the SM in our numerical calculations; this 
need not be the case experimentally\cite{me2}.)
In the case of $b\to s\gamma$ the required NLO real and virtual corrections 
to the operator matrix elements for the LRM are almost 
completely obtainable from the SM results when augmented by the new terms 
in the LO anomalous dimension matrix and through the use of 
left$\leftrightarrow$right symmetry. Additional terms arising 
from operators ${\cal O}_{12L,R}$ have yet to be included. Clearly we cannot 
claim to be performing a complete NLO treatment of the LRM case until 
all of the missing pieces have been calculated. However, this complete 
calculation is not necessary to demonstrate our points but certainly our 
analysis should be repeated once all the NLO pieces are in place for the LRM. 
The advantage of the present approach, however, is that when we turn off the 
effects associated with the various LRM contributions to both $b\to s\gamma$ 
and $b\to s\ell^+\ell^-$ we recover the usual NLO SM results.

\subsection{$b\to s\gamma$ in the LRM}

Using the central values of the quantities $m_t(m_t)=167$ GeV, $\mu=m_b$, 
$\alpha_s(M_Z)=0.118$, 
$m_c/m_b=0.29$, $\kappa=1$ and $B_l=0.1023$ we can calculate the rate for 
$b \to s\gamma$ in the LRM using the existing pieces of the NLO calculation 
that are presently available if we also provide sample values for the 
quantities $M_{W_2}$, $t_\phi$ and 
the relevant elements of the matrix $V_R$. Recall that we are looking here 
for particular non-decoupling regions in the LRM parameter space that 
essentially give the same result as the SM for the $b\to s\gamma$ branching 
fraction. (We are not interested in parameter space regions which differ only 
infinitesimally from the SM.) 
Let us first examine the 
simple scenario where $V_L=V_R$; this is the so-called manifest LRM. 
In this case, as discussed above, the $K_L-K_S$ mass difference and direct 
Tevatron collider searches 
require that $W_2$ be heavy; for purposes of demonstration we take 
$M_{W_2}=1.6$ TeV so that $t_\phi=\tan \phi$ is now the only free parameter 
as the $W_2$ contributions are now almost completely decoupled. 
Here we note that one of the interesting features uncovered by earlier 
analyses of $b\to s\gamma$ in the 
LRM{\cite {old}} was that left-right mixing terms associated with 
$t_\phi \neq 0$ can be enhanced by a helicity flip factor of $\sim m_t/m_b$ 
and can lead to significantly different predictions than the SM even in this
case where $V_L=V_R$ and $W_2$ is very heavy.

\vspace*{-0.5cm}
\nn
\begin{figure}[htbp]
\centerline{
\psfig{figure=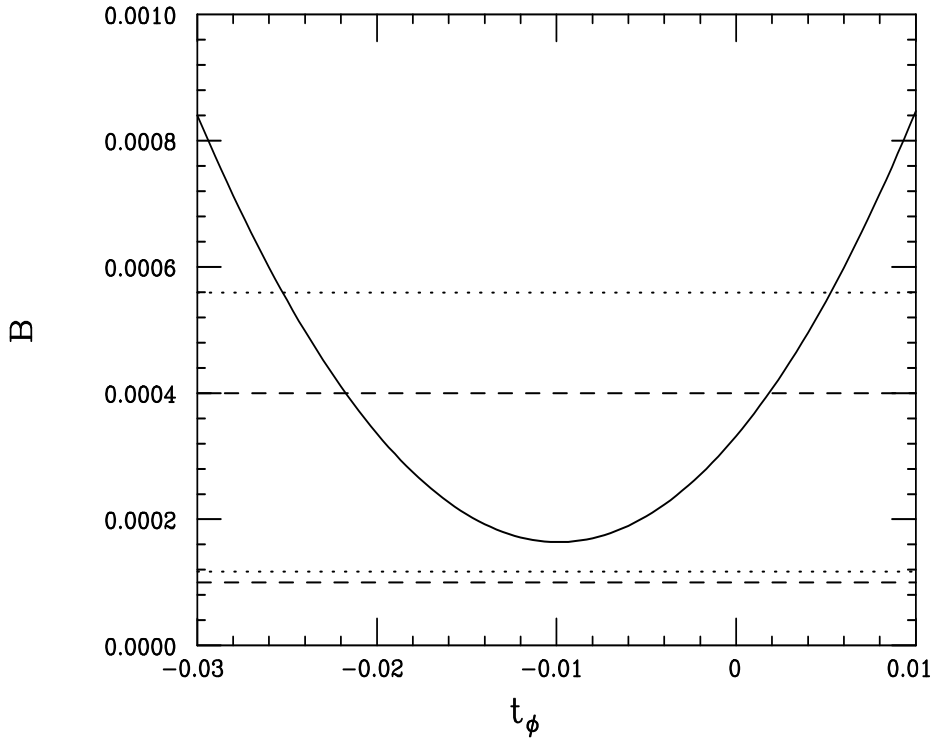,height=12.0cm,width=14.0cm,angle=-90}}
\vspace*{-0.9cm}
\caption{Prediction for the $b\to s\gamma$ branching fraction for the case 
$V_L=V_R$ and $M_{W_2}$=1.6 TeV as a function of $t_\phi$ in the LRM. 
The $95\%$ CL CLEO(ALEPH) allowed range lies inside the dashed(dotted) lines.}
\label{bsg}
\end{figure}
\vspace*{0.4mm}

Figure \ref{bsg} shows the prediction for the $b\to s\gamma$ branching fraction 
in this case and we see that the SM result is reproduced when $t_\phi=0$, as
expected, apart from a very small correction of order $M_{W_1}^2/M_{W_2}^2$. 
However we also see that a non-decoupling conspiratorial solution occurs when 
$t_\phi \simeq -0.02$, which yields a result which is exactly the same as the 
SM. From using $b \to s\gamma$ alone, the SM and this 
LRM case are indistinguishable, independent of what further improvements can 
be made in the branching fraction determination. This means that additional 
measurements would be necessary to separate these two cases. Of course, given 
present experimental data both the ALEPH and CLEO results allow for a rather 
broad range of $t_\phi$. 

\begin{figure}[htbp]
\centerline{
\psfig{figure=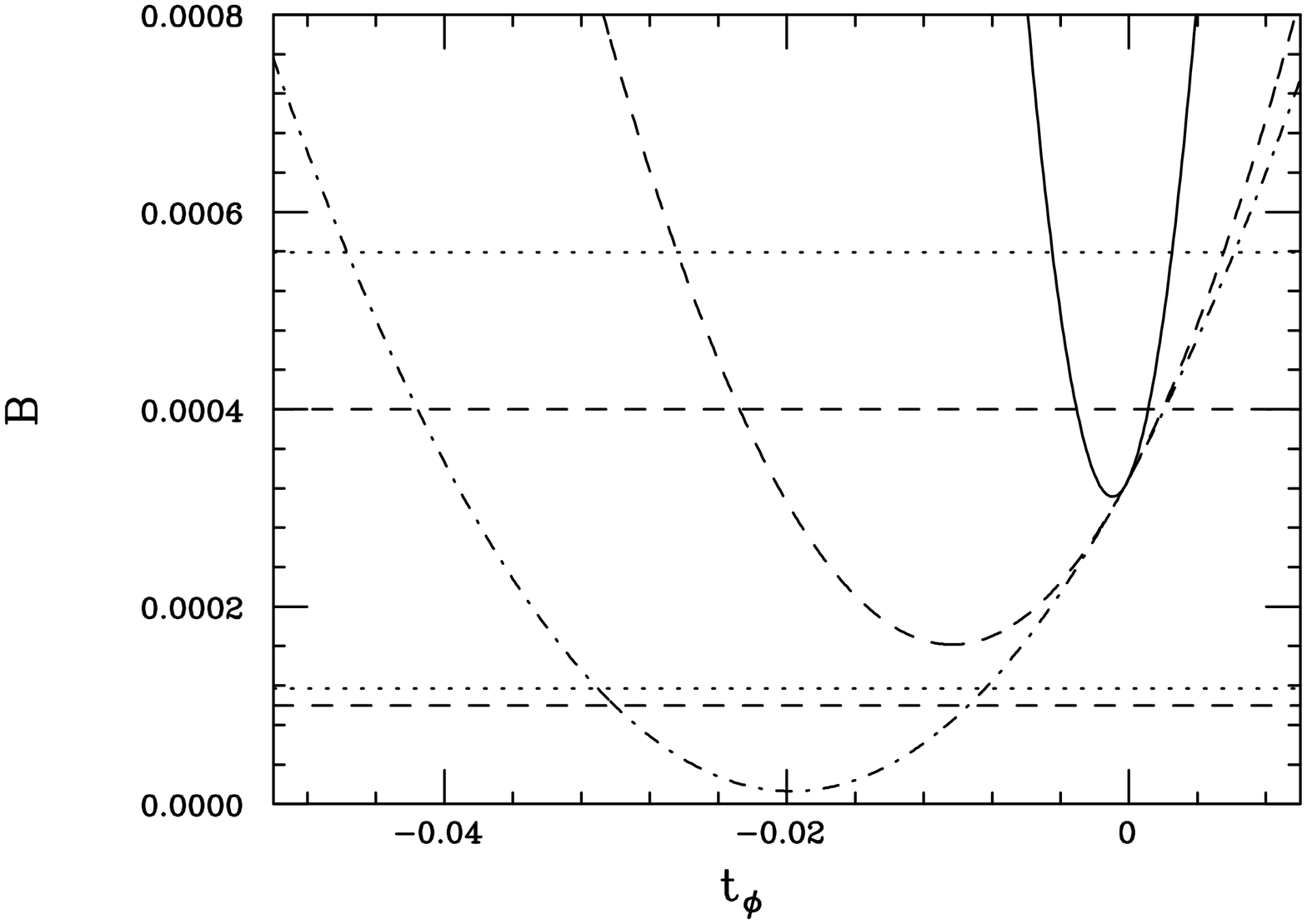,height=10.cm,width=12cm,angle=-90}}
\vspace*{-5mm}
\centerline{
\psfig{figure=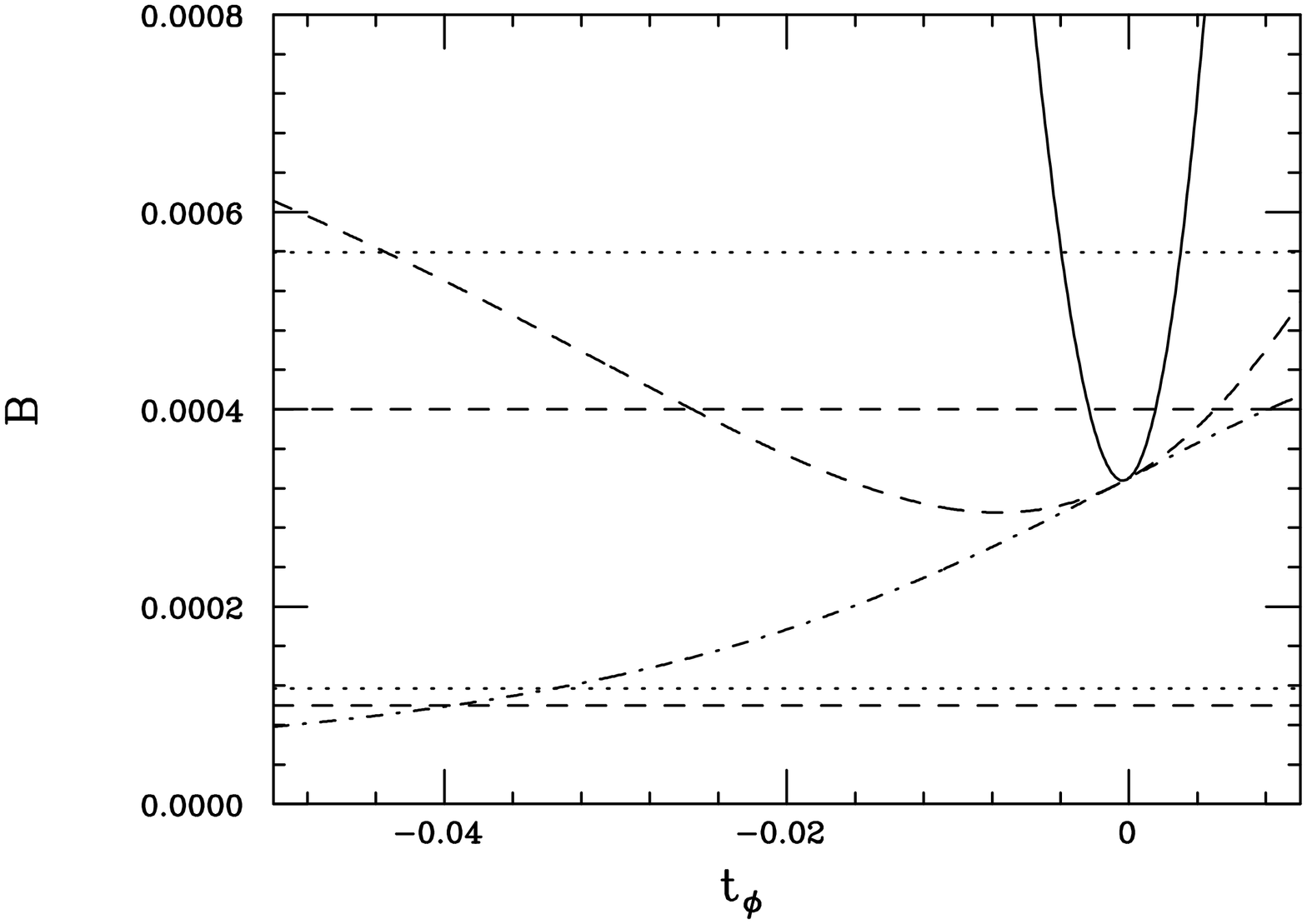,height=10.cm,width=12cm,angle=-90}}
\vspace*{-0.9cm}
\caption{Prediction for the $b\to s\gamma$ branching fraction for the cases 
$V_R=A(m)$ (top) and $V_R=D(n)$ (bottom) with $\kappa=1$ and $M_{W_2}$=0.8 TeV 
as a functions of $t_\phi$ in the LRM. 
The solid, dashed, dash-dotted curves correspond to $m,n$=1, 2 and 3, 
respectively. As before, the $95\%$ CL CLEO(ALEPH) allowed range lies inside 
horizontal the dashed(dotted) lines.}
\label{matd}
\end{figure}

We can now ask if corresponding conspiratorial regions exist for any of the 
more general forms of $V_R$ considered above. For example, if we calculate the 
$b\to s\gamma$ rate with the $A(m)$ and $D(n)$ matrices we obtain the results 
in Fig. \ref{matd}. For the $D(n)$ case, we see that only $n=2$ provides such
a region. For the other matrices we find conspiratorial regions when $m=1,2,3$ 
for matrix A and $p=2,3$ for matrix B; no such regions are 
found for matrices C, E and F. Note that $m=2$ in case A essentially 
corresponds to $V_L=V_R$. We denote these six conspiratorial cases as 
$V_L=V_R$, $A(1,3)$, 
$B(2,3)$ and $D(2)$ for later purposes. It is interesting to note that all 
of these cases lead to modest increases{\cite {alex}} 
in $B(b\to sg)$ by as much as $75\%$ above the SM prediction at LO. Of course, 
in addition to these conspiratorial solutions, we note from Fig. \ref{matd} 
that a rather wide range of $t_\phi$ is again allowed by current data from 
CLEO and ALEPH. 

\vspace*{-0.5cm}
\nn
\vspace*{0.4mm}

\subsection{$b\to s\ell^+\ell^-$ in the LRM}

For $b\to s \ell^+\ell^-$, the effective Hamiltonian leads to the matrix 
element (neglecting the strange quark mass but keeping the mass of the leptons) 
\begin{eqnarray}
{\cal M} & = &  {\sqrt 2 G_F\alpha\over\pi}\Bigg[ C_{9L}^{eff}
\bar s_L\gamma_\mu b_L\bar\ell\gamma^\mu\ell+C_{10L}\bar s_L\gamma_\mu
b_L\bar\ell\gamma^\mu\gamma_5\ell \nonumber \\
& & \quad\quad  -2C_{7L}^{eff} m_b\bar s_L i\sigma_{\mu\nu}{q^\nu\over q^2}
b_R\bar\ell\gamma^\mu\ell+L\to R\Bigg] \,,
\end{eqnarray}
where $q^2$ is the momentum transferred to the lepton pair. Note that 
$C_{9L,R}^{eff}$ contains the usual phenomenological long distance 
terms and that all the CKM elements are now contained in the 
coefficients themselves. From here we can directly 
obtain the expression for the double differential decay distribution 
\begin{eqnarray}
{dB \over {dz ds}} & = &B_l K  {3\alpha^2\over {16\pi^2}}
\beta (1-s)^2 \Bigg\{ \left[
(a_-^2+a_+^2)+(b_-^2+b_+^2)\right]{1\over {2}}\left[(1+s)-(1-s)\beta^2 z^2
\right] \nonumber \\
& & \quad\quad \left[(a_-^2-a_+^2)-(b_-^2-b_+^2)\right]\beta zs +4x(a_-a_++
b_-b_+) \nonumber \\
& & \quad\quad +{4\over {s^2}}(C_{7L}^2+C_{7R}^2)(1-s)^2(1-\beta^2 z^2) \\
& & \quad\quad -{2\over {s}}Re\left[C_{7L}(a_-+a_+)+C_{7R}(b_-+b_+)\right]
(1-s)(1-\beta^2 z^2) \Bigg\} \,, \nonumber
\end{eqnarray}
where $z=\cos \theta_{\ell^+ \ell^-}$, $s=q^2/m_b^2$, $x=m_\ell^2/m_b^2$, 
$\beta=\sqrt {1-4x/s}$, $a_\pm=C_{9L}^{eff}\pm C_{10L}+2C_{7L}/s$ and 
$b_\pm=a_\pm$ with the replacement $L\to R$. As usual $\theta_{\ell^+ \ell^-}$ 
is the angle between the lepton direction and that of the original $b$ quark 
in the lepton pair center of mass frame. Multiplication of the above 
expression by the total number of produced $B$ mesons gives the double 
differential event distribution.
We normalize this rate to the usual semileptonic 
branching fraction ($B_l=0.1023$), including finite $m_c/m_b=0.29$ and QCD 
corrections with $\alpha_s(M_Z)=0.118$, which are fully included in the 
overall normalization parameter 
$K$ (see Ali \etal, in Ref. \cite{jlh}). LRM corrections to the semileptonic 
rate are, of course, also included; here the assumption that the mass of the
right-handed neutrino $m_N>m_b$ 
becomes relevant.  For numerical purposes we take $m_N=250$ GeV in the
calculation of the right-handed box diagram for $b\to s\ell^+\ell^-$, but the 
precise value of $m_N$ is not important for our purposes of demonstration. 

\begin{figure}[htbp]
\centerline{
\psfig{figure=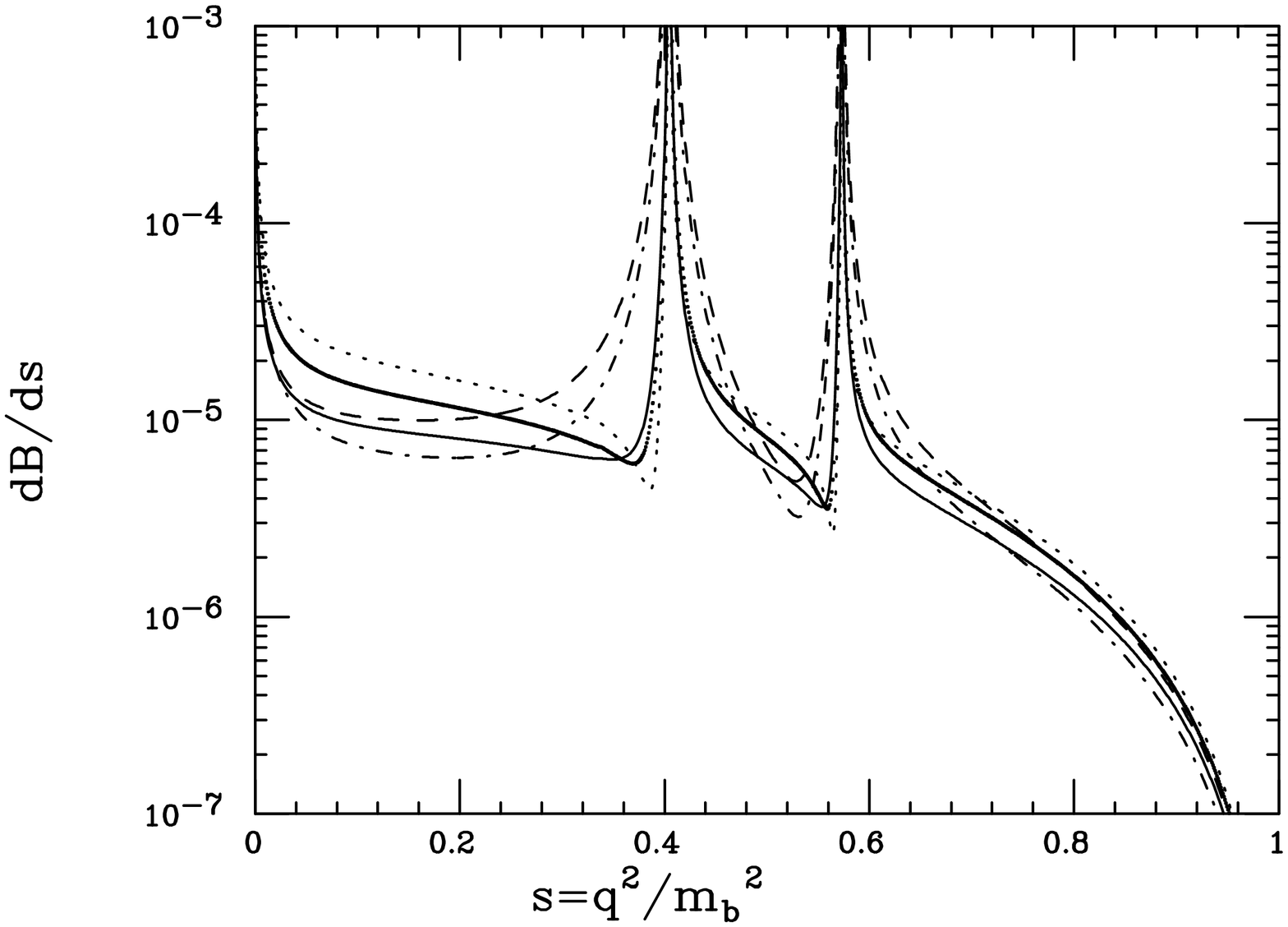,height=9.1cm,width=9.1cm,angle=-90}
\hspace*{-5mm}
\psfig{figure=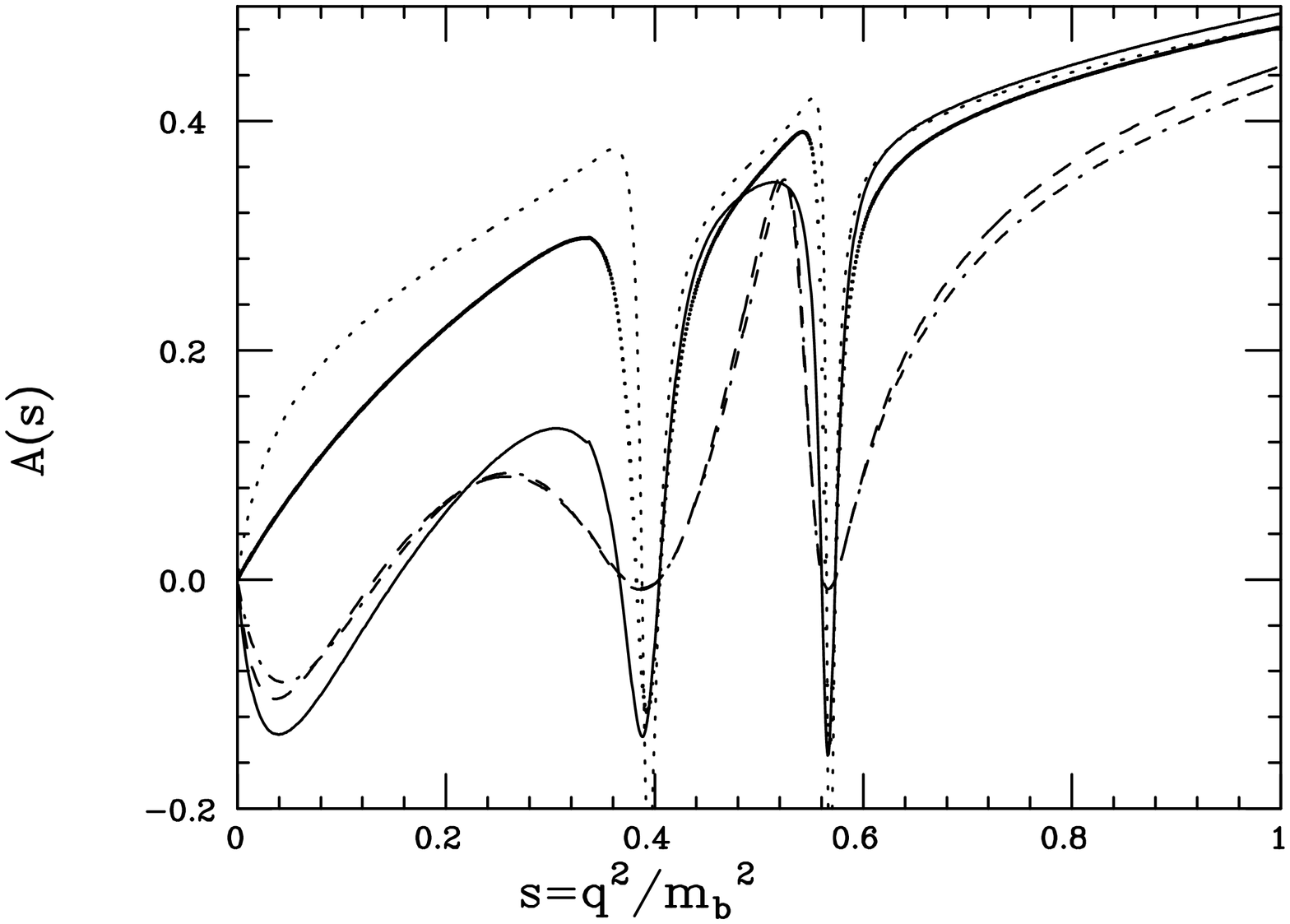,height=9.1cm,width=9.1cm,angle=-90}}
\vspace*{-0.6cm}
\caption{Differential decay distribution and lepton forward-backward 
asymmetry for the decay 
$b\to s\ell^+\ell^-$ in the SM(solid) and four representative models in the  
LRM parameter space which yield the SM value for the $b\to s\gamma$ branching 
fraction and satisfy all other existing experimental constraints: 
$V_L=V_R$(square-dot), $A(1)$(dash), $A(3)$(dot) and $D(2)$(dash-dot).} 
\label{bsll}
\end{figure}

From this double differential distribution we can compute both the lepton 
pair invariant mass distribution by integration over $z$ as well as the 
leptonic forward-backward asymmetry.  
The asymmetry is given by the expression 
\begin{equation}
A(s)= {{\int_0^1 {dB \over {ds dz}} \, dz -\int_{-1}^0 {dB \over 
{ds dz}} \, dz}\over {\int_{-1}^1 {dB \over {ds dz}} \, dz}}  \,.
\end{equation}
These two observables are shown in Fig. \ref{bsll} for the SM as 
well as for four of the LRM conspiratorial 
cases discussed above, assuming massless 
leptons in the final state. As can be easily seen here the predictions of the 
four LRM cases are quite distinct from those of the SM even though they all 
yield the same prediction for $B(b\to s\gamma)$. 

\vspace*{-0.5cm}
\nn
\vspace*{0.1mm}

Other observables, such as the longitudinal polarization asymmetry 
of the $\tau$'s in $b\to s\tau^+\tau^-$, can be obtained in a straightforward 
fashion from the expressions above and an 
$L\to R$ augmentation of the expressions provided by Hewett{\cite {jlh}}. 
These are the only $b\to s\ell^+\ell^-$ observables we will make use of in the 
analysis below; the possibility of employing{\cite {corr3}} the transverse 
polarization of the $\tau$'s has been neglected.  In a similar spirit we ignore 
the possible information that one could gain from future photon polarization 
measurements in $b\to s\gamma$.

\section{Analysis Results}

The essential aspects of our procedure can be found in the work of 
Hewett{\cite {jlh}} and we closely follow the discussion of this author. 
The analysis makes use of the following observables. 
For the process $b\to s\ell^+\ell^-$, 
we consider the lepton pair invariant mass distribution 
described by $dB/ds$, 
and the lepton pair forward-backward asymmetry, $A(s)$,  for $\ell=e,\mu$, 
and 
$\tau$, as well as the tau longitudinal polarization asymmetry $P_\tau(s)$. 
We will neglect the $\mu$ mass for simplicity and directly combine the 
$e$ and $\mu$ samples. We also include $B(b\to s\gamma)$ to this list of 
observables. The lepton pair invariant mass 
spectrum is divided into 9 bins which are distributed as follows: 
6 bins of equal size,  $\Delta s =0.05$, are 
taken in the low dilepton mass region below the $J/\psi$
resonance, $0.02\leq s \leq 0.32$, and 3 bins are in the high 
dilepton mass region above the $\psi'$ pole and are taken to be 
$0.6\leq s \leq 0.7$, $0.7\leq s \leq 0.8$, and $0.8\leq s \leq 1.0$. By 
using this set of bins we completely avoid the regimes where both long distance 
and resonance contributions are clearly important. 

Our analysis proceeds as follows. For a given conspiratorial 
choice of $V_R$ derived above we use Monte Carlo techniques to generate 
binned ``data'' associated with the above quantities.
For the $b\to s\ell^+\ell^-$ observables 
we assume that the errors will remain statistically dominated, while for 
$b\to s\gamma$ we assume a purely 
systematically dominated error of $7\%$ arising from 
both experimental and theoretical uncertainties. These distributions are 
then generated for an integrated
luminosity of either $5\times 10^7$ or $5\times 10^8$ $B\bar B$ pairs; these 
correspond to the expected
total luminosity of a couple of years of running at future $B$-factories 
on the $\Upsilon(4S)$ and at the LHC, respectively.  
For $b\to s\ell^+\ell^-$ the number of events per bin is given by
\begin{equation}
N_{\rm bin}=\lum \int_{s_{\rm min}}^{s_{\rm max}} {dB \over
ds} \, ds \,,
\end{equation}
and the integrated average value of the asymmetries for each bin is then
\begin{equation}
\langle A\rangle_{\rm bin} ={\lum\over N_{\rm bin}}
\int_{s_{\rm min}}^{s_{\rm max}} A(s) \, {dB \over ds} \, ds \,,
\end{equation}
where $dB/ds$ can be obtained from the double differential expression above.

\vspace*{-0.5cm}
\nn
\begin{figure}[htbp]
\centerline{
\psfig{figure=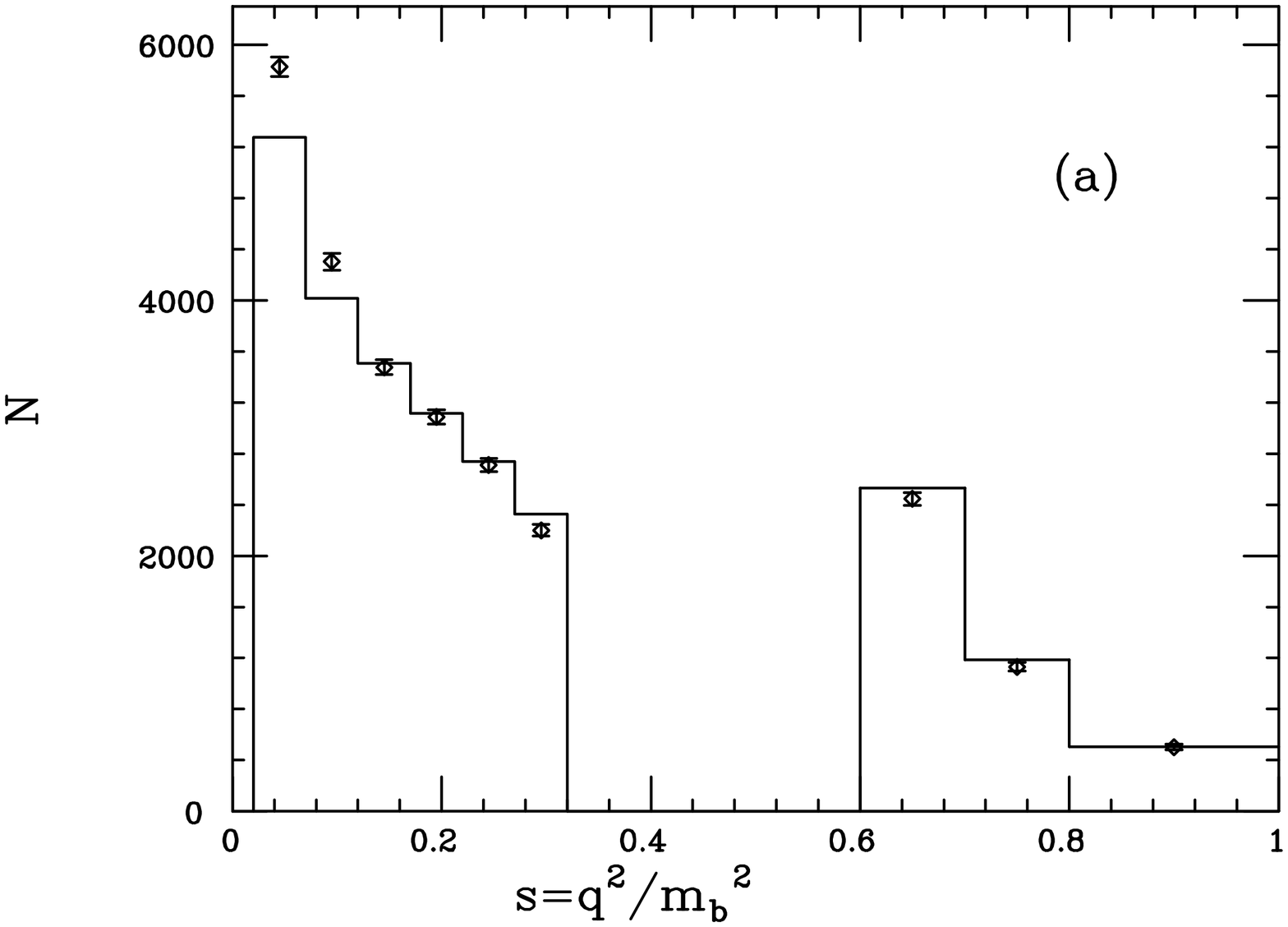,height=9.1cm,width=9.1cm,angle=-90}
\hspace*{-5mm}
\psfig{figure=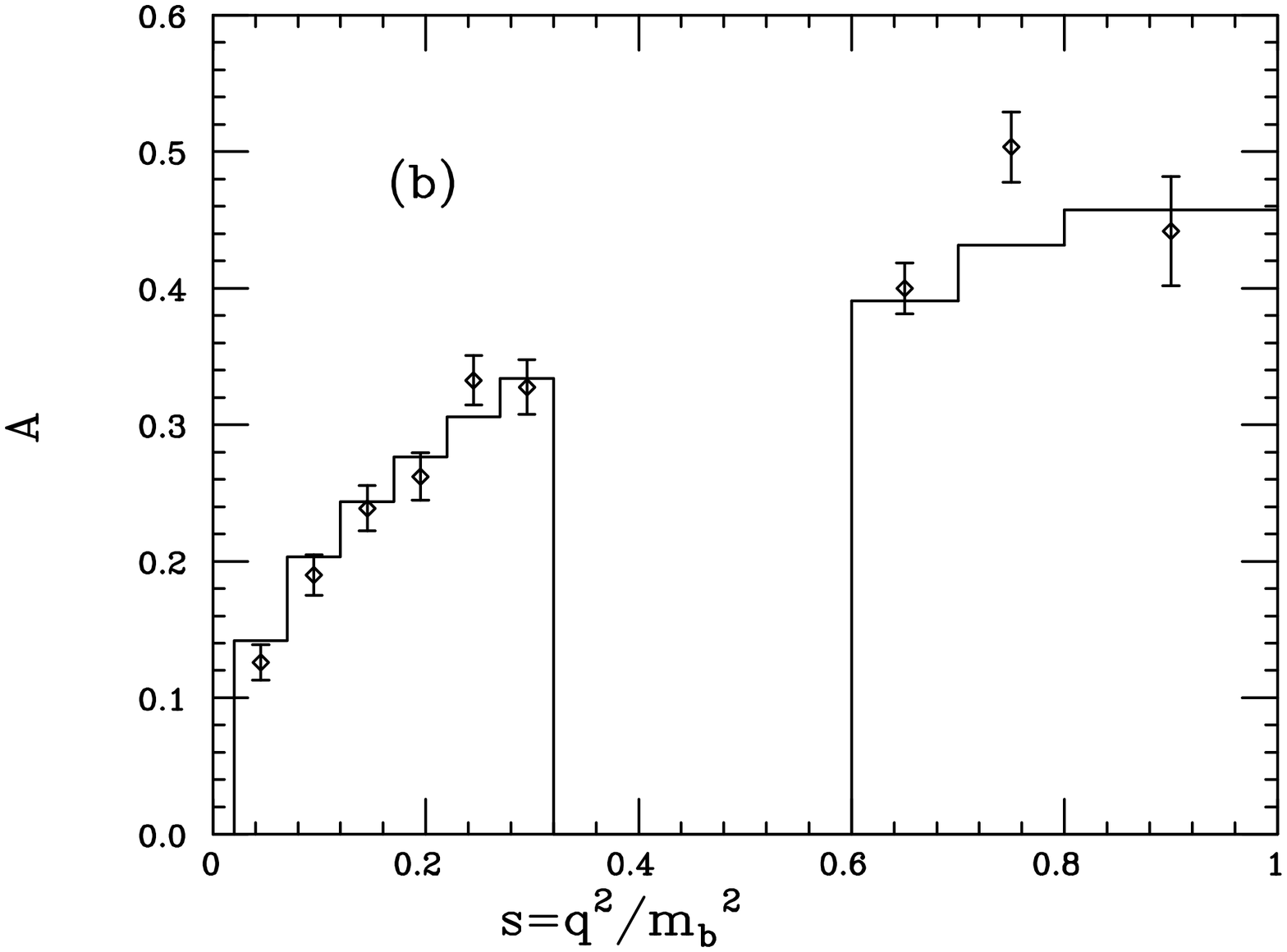,height=9.1cm,width=9.1cm,angle=-90}}
\vspace*{-0.6cm}
\caption{Comparison of a typical Monte Carlo event sample generated with the 
right-handed 
mixing matrix $B(3)$ for (a) the number of events per bin and (b)
the bin averaged asymmetry. $e$ and $\mu$ final states are 
summed and the histogram represents the best fit to the data. A sample of 
$5\times 10^8$ $B\bar B$ pairs has been assumed.} 
\label{fit}
\end{figure}
\vspace*{0.1mm}

Once the data is generated for each model we then perform a three dimensional
$\chi^2$ fit assuming only the usual three coefficients are present, \ie, 
$C_{7L,9L,10L}(\mu)$.  This is done according to the usual prescription
\begin{equation}
\chi^2_i=\sum_{\rm bins} \left( {Q_i^{\rm obs}-Q_i(C_L)\over \delta Q_i}
\right)^2 \,,
\end{equation}
where $Q_i^{\rm obs}$, $Q_i(C_L)$, $\delta Q_i$ represent the ``data'', the 
result of the expectations for a given set of $C_{7L,9L,10L}(\mu)$ values, 
and the error for each observable quantity $Q_i$.  The $C_{7L,9L,10L}$ 
are then varied until a $\chi^2$ minimum is obtained. 
Note that there are 27 bins 
of data in addition to $B(b\to s\gamma)$; allowing for the three free 
parameters $C_{7L,9L,10L}$ means that the fit has 25=27+1-3 degrees of 
freedom. If the SM were realized, 
we would expect such a fit to yield the SM values of $C_{7L,9L,10L}$,
within errors, with a good $\chi^2\simeq 25$ as was shown by 
Hewett{\cite {jlh}}. If new physics is present but the operator basis is not 
extended we would again expect a comparably good $\chi^2$ fit at values of 
$C_{7L,9L,10L}$ which would now exclude the SM at some confidence level. 
In our case with an extended operator basis we would hope that the best fit 
obtained by varying these three coefficients alone is not very good 
thus demonstrating that the three parameter fit is insufficient.

\vspace*{-0.5cm}
\nn
\begin{figure}[htbp]
\centerline{
\psfig{figure=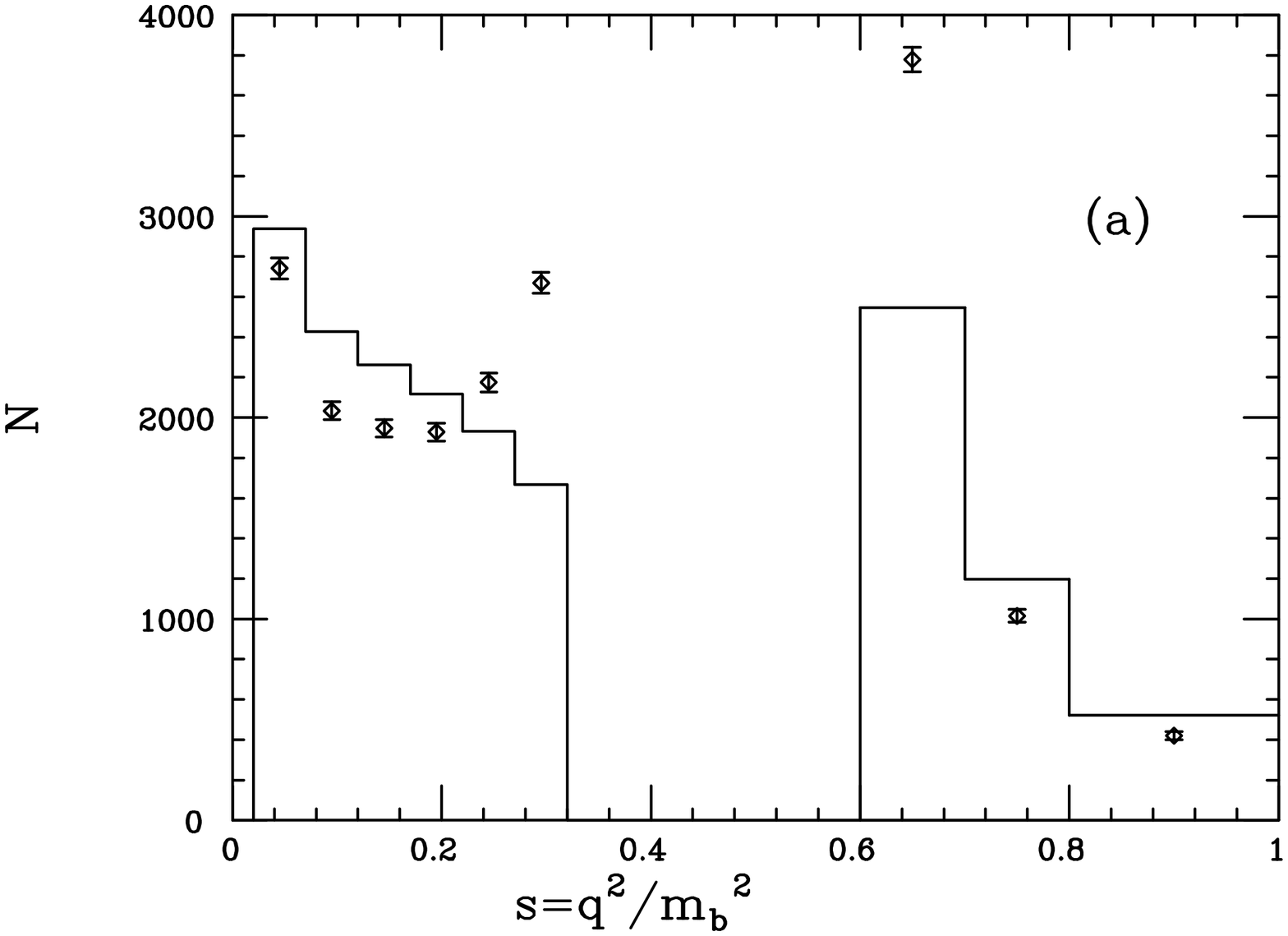,height=9.1cm,width=9.1cm,angle=-90}
\hspace*{-5mm}
\psfig{figure=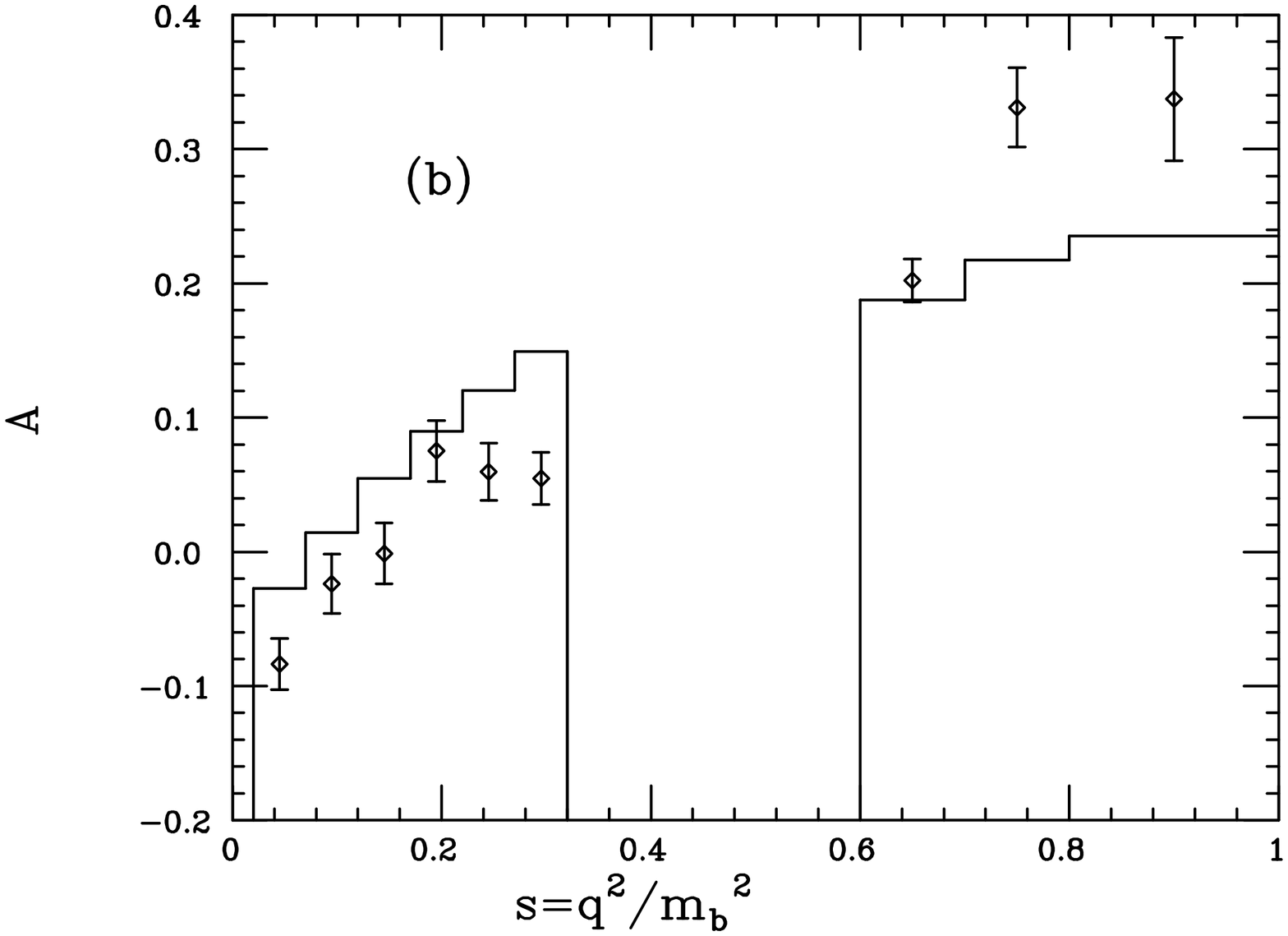,height=9.1cm,width=9.1cm,angle=-90}}
\vspace*{-0.6cm}
\caption{Same as the previous figure but now for matrix A(1).} 
\label{fita}
\end{figure}
\vspace*{0.1mm}

We now give a few examples of this type of analysis.  We first consider a 
typical Monte Carlo data sample generated for the matrix $B(3)$, assuming 
$5\times 10^8$ $B\bar B$ pairs are produced, and examine the results of the 
best fit.  This is shown in Fig. \ref{fit} for the two observables $N$ and $A$ 
for combined $e$ and $\mu$ data samples. Note that in the low $s$ bins the best 
fit underestimates (overestimates) the number of events at the low (high) 
energy end and generally underestimates the asymmetry. 
Combining all of the observables this particular ``data'' set leads 
to a $\chi^2/d.o.f$ of 187.1/25. We note that for 25 $d.o.f.$, $\chi^2$ values 
of 37.65(44.31,52.61,60.14) 
correspond to probabilities of a consistent fit of $5(1,0.1,0.01)\%$, 
respectively. Thus it 
is clear that for this particular sample the ``data'' are not consistent 
with the assumption of only three active operators and that an extended 
operator basis is required. 

\begin{figure}[htbp]
\centerline{
\psfig{figure=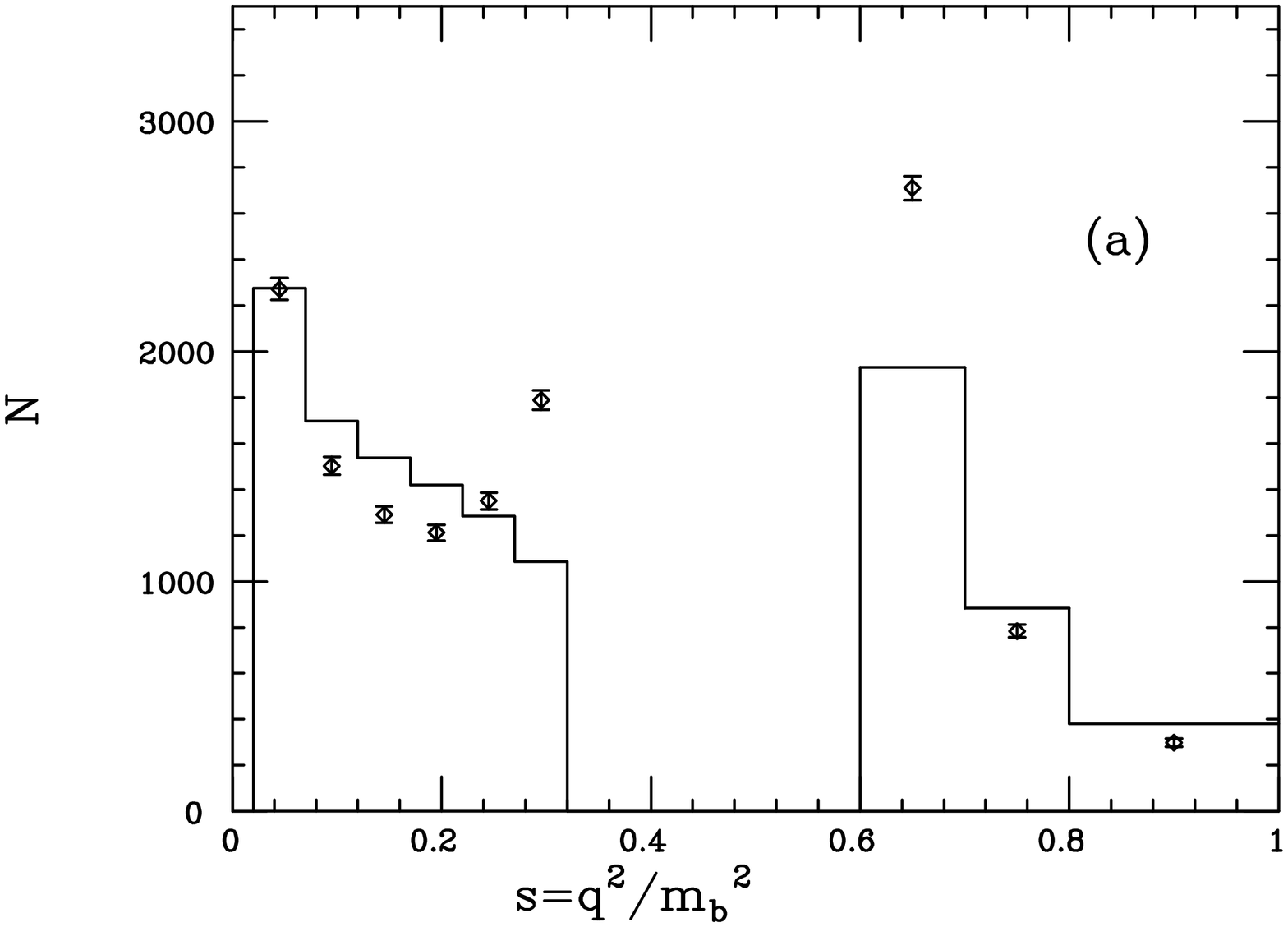,height=9.1cm,width=9.1cm,angle=-90}
\hspace*{-5mm}
\psfig{figure=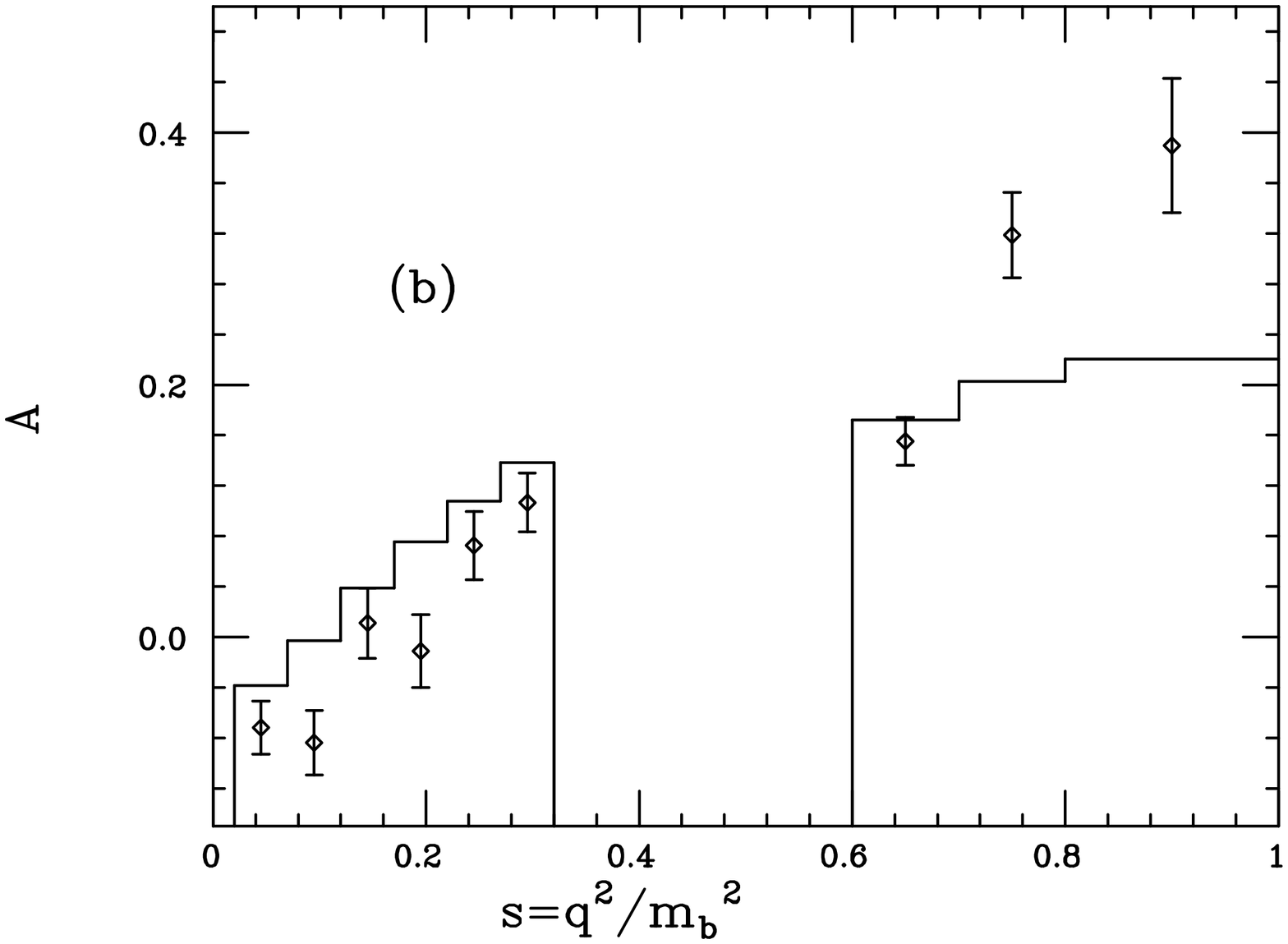,height=9.1cm,width=9.1cm,angle=-90}}
\vspace*{-0.6cm}
\caption{Same as the previous figure but now for matrix D(2).}  
\label{fitd}
\end{figure}

Figures \ref{fita} and \ref{fitd} show examples of 
generated data and the corresponding best three parameter fits for the 
right-handed matrices $A(3)$ and $D(2)$. In both these cases the 
best fits are simply incapable of obtaining the correct shape presented by the 
data. For these two cases the $\chi^2/d.o.f.$ from these fits 
are found to be 1187.2/25 and 764.1/25, respectively, for 
a sample of $5\times 10^8$ $B\bar B$ pairs. Thus we again see that these 
particular data sets are not consistent with the three Wilson coefficient fit 
hypothesis. 

Of course, to really ascertain if the new physics of the extended operator 
basis should be visible we need to generate many sets of data for each of the 
$V_R$ assumptions and determine the fraction of the time that the resulting 
$\chi^2$ values exceed those listed above 
for each of the fixed probabilities. To be specific we generate 1000 sets of 
data for each of the models above and perform the $\chi^2$ fitting 
procedure for each data set. 
For very large data samples, \ie, for $5\times 10^8$ $B\bar B$ pairs, we find 
that for all of the above choices of $V_R$ we obtain fit probabilities below 
$0.01\%$ almost $100\%$ of the time. This means that it would be quite 
clear in this case that an extended operator basis is required. 
For smaller data samples, \ie, for 
$5\times 10^7$ $B\bar B$ pairs, the results are much more model dependent and 
are given in Table 2. It seems that $V_L=V_R$ represents the worst case 
scenario. (To test the stability of our results we generated 10000 sets of 
data for this case and confirmed the results shown in the Table.) Obviously
it is quite important to decide at what 
level of probability one is willing to exclude the three operator fit before 
drawing any conclusions about an extended operator basis. However, it is clear 
that even for very low probabilities, $\sim 0.01\%$, we see that the three 
parameter fit will fail on average a reasonable fraction, $\sim 45\%$, of 
the time.

\vspace*{-0.5cm}
\nn
\vspace*{0.1mm}
%

%
\begin{table*}[htpb]
\leavevmode
\begin{center}
\label{results}
\begin{tabular}{lcccc}
\hline
\hline
Matrix & $P=5\%$ & $P=1\%$ & $P=0.1\%$ & $P=0.01\%$ \\
\hline
$V_L=V_R$ & $\sim 39\%$ & $\sim 23\%$ & $\sim 11\%$ & $\sim 6\%$ \\
$A(1)$     & $\sim 100\%$ & $\sim 100\%$ & $\sim 100\%$ & $\sim 100\%$ \\
$A(3)$     & $\sim 90\%$ & $\sim 76\%$ & $\sim 51\%$ & $\sim 30\%$ \\
$B(2)$   & $\sim 47\%$ & $\sim 28\%$ & $\sim 15\%$ & $\sim 8\%$ \\
$B(3)$   & $\sim 92\%$ & $\sim 80\%$ & $\sim 56\%$ & $\sim 36\%$ \\
$D(2)$    & $\sim 100\%$ & $\sim 100\%$ & $\sim 100\%$ & $\sim 99\%$ \\
\hline
\hline
\end{tabular}
\caption{Fraction of the time the effects of the extended operator basis are 
observable, with different choices of the consistency of fit probabilities, 
for each of the $V_R$ choices described in the text assuming a 
sample of $5\times 10^7$ $B\bar B$ pairs. 
These results are based on 1000 Monte Carlo data samples 
for each of the models.}
\end{center}
\end{table*}

\section{Discussion and Conclusions}

The inclusive rare decays $b\to s\gamma$ and $b\to s\ell^+\ell^-$ have been 
and will continue to be subjects of intense interest since they offer 
unique opportunities to probe for new physics beyond the Standard Model. Both 
these decay modes are quite clean theoretically and future $B$-factories will 
produce large event samples in both cases, allowing for in-depth studies of 
their associated observables.  As we know, in the SM and in
many of its extensions these decay observables can be 
expressed in terms of only three {\it a priori} unknown parameters,
corresponding to the values of the Wilson coefficients of three operators 
at the scale $\mu \sim m_b$. The global fit approach 
provides the best model independent technique for obtaining the values of 
these coefficients at the low scale which can then be compared with the 
expectations of a given model. 
In this paper we have demonstrated, using the Left-Right Symmetric Model as an 
example, that with the statistics available at 
future $B$-factories it will be possible to observe the rather unique 
situation where this global fit to the canonical three coefficients fails. 
This result would tell us that not only does new physics exist beyond the SM 
but that this new physics {\it requires} an extended operator basis.

\noindent{\Large\bf Acknowledgements}

The author would like to thank J.L. Hewett, A. Masiero, G. Isidori, J. Wells, 
M. Worah and A. Kagan for discussions related to this work.

\newpage

%
\def\MPL #1 #2 #3 {Mod. Phys. Lett. {\bf#1},\ #2 (#3)}
\def\NPB #1 #2 #3 {Nucl. Phys. {\bf#1},\ #2 (#3)}
\def\PLB #1 #2 #3 {Phys. Lett. {\bf#1},\ #2 (#3)}
\def\PR #1 #2 #3 {Phys. Rep. {\bf#1},\ #2 (#3)}
\def\PRD #1 #2 #3 {Phys. Rev. {\bf#1},\ #2 (#3)}
\def\PRL #1 #2 #3 {Phys. Rev. Lett. {\bf#1},\ #2 (#3)}
\def\RMP #1 #2 #3 {Rev. Mod. Phys. {\bf#1},\ #2 (#3)}
\def\ZPC #1 #2 #3 {Z. Phys. {\bf#1},\ #2 (#3)}
\def\IJMP #1 #2 #3 {Int. J. Mod. Phys. {\bf#1},\ #2 (#3)}


\begin{thebibliography}{99}
\bibitem{joa}
The author would like to thank JoAnne Hewett for suggesting this analogy.
\end{thebibliography}
\end{document}